\documentclass[onecolumn,draftclsnofoot,12pt]{IEEEtran}
\usepackage{amssymb}
\usepackage{amsfonts}

\usepackage{enumerate}

\usepackage{enumerate}
\usepackage{amsmath,amsthm}
\usepackage{mathtools}
\usepackage{algorithm}
\usepackage[noend]{algpseudocode}
\usepackage{float}
\usepackage{hyperref}
\usepackage{color}
\usepackage{makeidx}
\usepackage{bbm}
\usepackage{graphicx}
\usepackage{lipsum}
\usepackage{soul}
\usepackage{tabularx}
\usepackage{dsfont}
\usepackage[table,xcdraw]{xcolor}

\usepackage{amsfonts}
\usepackage{times}
\usepackage{graphicx}
\usepackage{latexsym}
\usepackage{dsfont}
\usepackage{amssymb}
\usepackage{amsmath}
\usepackage{cite}
\usepackage{verbatim}
\usepackage{subfigure}

\def\bb0{{\mathbb{0}}}

\def\bb{{\mathbf{b}}}

\def\bp{{\mathbf{p}}}

\def\b0{{\mathbf{0}}}

\def\sf0{{\mathsf{0}}}

\usepackage{epstopdf}

\newcommand{\sref}[1]{{Section}~\ref{#1}}
\newcommand{\fref}[1]{{Fig.}~\ref{#1}}

\algnewcommand{\Initialize}[1]{%
	\State \textbf{Initialization:} \parbox[t]{.8\linewidth}{\raggedright #1}}

\newcommand{\subto}{\operatorname{s.t.}}
\newcommand{\argmin}{\operatornamewithlimits{arg\min}}

\begin{document}
\title{Deep Learning of Near Field Beam Focusing in Terahertz Wideband Massive MIMO Systems}
\author{Yu Zhang and Ahmed Alkhateeb \thanks{Yu Zhang and Ahmed Alkhateeb are with Arizona State University (Email: y.zhang, alkhateeb@asu.edu). This work is supported by the National Science Foundation under Grant No. 1923676.}}
\maketitle

\begin{abstract}

Employing large antenna arrays and utilizing large bandwidth have the potential of bringing very high data rates to future wireless communication systems. However, this brings the system into the near-field regime and also makes the conventional transceiver architectures suffer from the wideband effects. To address these problems, in this paper, we propose a low-complexity frequency-aware beamforming solution that is designed for hybrid time-delay and phase-shifter based RF architectures. To reduce the complexity, the joint design problem of the time delays and phase shifts is decomposed into two subproblems, where a signal model inspired online learning framework is proposed to learn the shifts of the quantized analog phase shifters, and a low-complexity geometry-assisted method is leveraged to configure the delay settings of the time-delay units. Simulation results highlight the efficacy of the proposed solution in achieving robust performance across a wide frequency range for large antenna array systems.

\end{abstract}

\section{Introduction} \label{intro}


Employing large antenna arrays and utilizing the large bandwidth available at millimeter wave and terahertz (THz) bands are two key trends in current and future wireless communication systems.
However, these new characteristics also bring new design challenges.
Two particular challenges are: (i)  With a large antenna array aperture, the system is more likely to operate in the near-field region. This makes the conventional far-field planar wave propagation assumption invalid and challenges the beamforming design problems. (ii) With the wide bandwidth, the classical phase shifter based analog-only beamforming experiences beam squinting/mis-focusing, leading to degradation in the beamforming gain and increase in the introduced network interference.
These problems motivate adopting new transceiver architectures  and developing novel beamforming solutions for near-field wideband communication systems, which is the focus of this paper.

\textbf{Prior Work:}
A common approach to improve the performance of a wideband system is by equipping the antenna arrays with the time delay (TD) units which can introduce a constant group delay to the different frequency components of the transmit/receive signals. For instance, in \cite{Longbrake2012,Ghaderi2019,Boljanovic2021}, the authors demonstrate the potential of such TD units based arrays in achieving robust beamforming gain performance and in facilitating other operations (such as beam training and alignment) in the wideband systems. However, the TD-based arrays come with a higher manufacture cost and they also consume more power than the conventional phase-shifter (PS) based antenna arrays.
With the motivation to address this increased cost and power consumption, \cite{Liu2019Space-Time,Myers2022} studied the near-field wideband beam focusing problem using PS-only architectures. Although the average performance within the considered bandwidth is optimized, the peak gains are still majorly sacrificed.
To strike a balance between performance and hardware cost/power consumption, the hybrid TD-PS based architecture is introduced, where the wideband beamforming problem is studied under both far-field \cite{Dai2022} and near-field \cite{Cui2021Near-Field} channel models.
However, the achieved performance in the existing work relies on accurate channel knowledge, which incurs high training and acquisition overhead given the large number of antennas.

\textbf{Contribution:}
In this paper, we develop a low-complexity beamforming approach for TD-PS architecture that is \textbf{capable of learning and optimizing the near-field wideband beams without requiring any channel knowledge.}
Specifically, an online learning framework that is inspired by the signal model is developed to learn the phase shifts of the quantized analog PSs and achieve fast convergence despite the large numbers of antennas.
Further, a low-complexity geometry-assisted method is devised to configure the delay settings of the TD units.
The proposed framework relies only on the power measurements of the received signal at different subcarriers, and does not require any explicit channel or location information.
Simulation results demonstrate the capability of the proposed solution in learning optimized near-field beams that achieve robust performance in practical wideband large-antenna array systems.

\section{System and Channel Models} \label{sec:System}

\subsection{System Model}

We consider a system where a THz massive MIMO base station (BS) with $M$ antennas is communicating with a single-antenna user equipment (UE).
The system adopts OFDM transmission/reception with $K$ subcarriers and operates at a center frequency $f_c$ over a bandwidth $B$.
In the uplink, if the UE transmits a symbol $s_k\in\mathbb{C}$ at the $k$-th subcarrier, then the received signal at the BS after combining can be expressed as
\begin{equation}\label{rec-signal}
  y_k = {\bf w}^H{\bf h}_ks_k + {\bf w}^H{\bf n}_k, ~ \forall k=1,\dots,K,
\end{equation}
where we assume that the transmitted symbol $s_k$ satisfies the average power constraint $\mathbb{E}\{|s_k|^2\}=\frac{P_\mathrm{T}}{K}$ with $P_\mathrm{T}$ denoting the total transmit power.
${\bf w}$ is the combining vector, ${\bf h}_k\in\mathbb{C}^{M\times 1}$ is the uplink channel vector between the UE and the BS at the $k$-th subcarrier, and ${\bf n}_k \sim \mathcal{CN}(0, \sigma_k^2{\bf I})$ is the receive noise vector at the BS with $\sigma_k^2$ being the noise power.

\begin{figure}[t]
	\centering
	\includegraphics[width=.85\columnwidth]{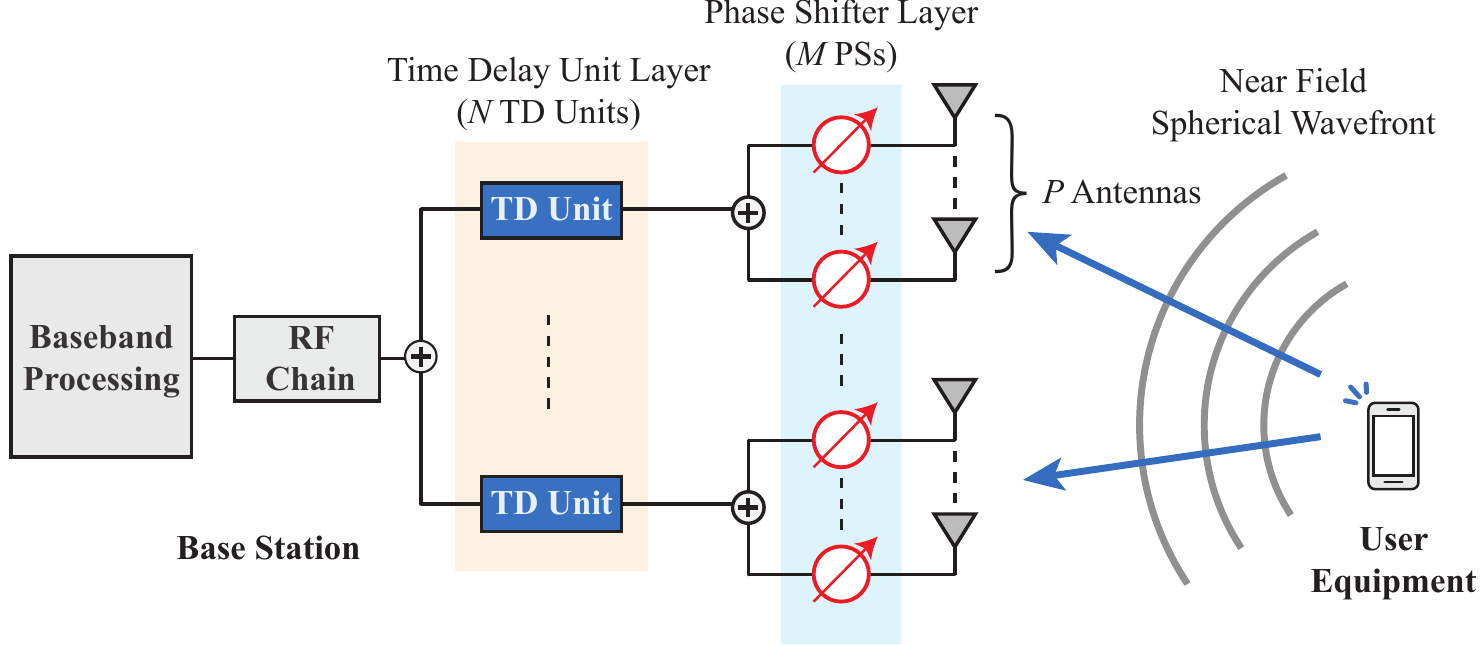}
	\caption{The considered system architecture where we assume a general non-uniform linear array that adopts analog PSs and TD units. The user equipment is assumed to be within the near field region of the large-dimensional array.}
	\label{sys-ps-ttd}
\end{figure}

The combining vector ${\bf w}$ in \eqref{rec-signal} takes different forms in different transceiver architectures.
In this paper, we consider a practical system where the BS has only one RF chain and employs analog-only beamforming using a network of $r$-bit quantized PSs.
Further, to mitigate the severe beam split effect in the wideband systems, we assume that the system is equipped with time-delay units.
These TD units are RF circuit blocks that introduce a constant group delay to the different frequency components of the received signal \cite{Boljanovic2021}.
For example, if the $n$-th TD unit is configured to introduce a delay of $\tau_n$ seconds, then the phase shift incurred for the signal at the $k$-th subcarrier is $-2\pi f_k\tau_n$.
As shown in \fref{sys-ps-ttd}, we assume that every $P$ PSs are connected to one TD unit, i.e., forming a partially connected architecture.
Without loss of generality, we assume that the system adopts a total number of $N$ TD units such that $M=NP$.
With this architecture, and unlike the conventional PS-only beamforming, the beamformer becomes frequency-dependent.
Specifically, the effective combining vector at the frequency $f_k$ is given by
\begin{equation}\label{bf-vec}
  {\bf w}_k = \frac{1}{\sqrt{M}}e^{j\left(-2\pi f_k\boldsymbol{\tau}\otimes\mathbf{1}_P + \boldsymbol{\theta}\right)},
\end{equation}
where $\otimes$ is Kronecker product, $\boldsymbol{\tau}\in\mathbb{R}^{N\times 1}$ is the delay vector, $\boldsymbol{\theta}\in\mathbb{R}^{M\times 1}$ is the phase vector (of the phase shifters), and $\mathbf{1}_P$ denotes the $P\times 1$ all one vector.
{
In the rest of the paper, we will use the approximation
\begin{equation}
  {\bf w}(f) \approx {\bf w}_k, ~ \forall f\in\left[f_k-\frac{B}{2K}, f_k+\frac{B}{2K}\right],
\end{equation}
i.e., we will assume that the effective beamforming is constant across each subcarrier.
}

\subsection{Near Field Wideband Channel Model} \label{subsec:H}

Without loss of generality, we assume a 2-D geometry, where both the BS antenna array and the UE's antenna are at the x-y plane.
The BS is assumed to employ a linear array that is aligned with the y axis and that has a half aperture of $\frac{D}{2}$.
The center of the BS array is assumed to be at the origin. Based on that, the positions of the $M$ BS antenna elements can be written as ${\bf p}_m=[0, \frac{D}{2}\alpha_m]^T, \forall m=1,2,\dots,M$, where $\alpha_m$ takes a value in $[-1, 1]$, i.e.,
\begin{equation}
  -1 \le \alpha_M < \alpha_{M-1} < \cdots < \alpha_2 < \alpha_1 \le 1.
\end{equation}
It is worth noting that we do not assume that the BS adopts a uniform linear array (ULA), meaning that $\alpha_m, \forall m$ can be any values in $[-1, 1]$ and only need to satisfy the sequential constraint, i.e., $\alpha_{M} < \cdots < \alpha_2 < \alpha_1$.
Finally, we denote the UE's position as ${\bf q}=[q_x, q_y]^T\in\mathbb{R}^2$.
Hence, the distance between the UE's antenna and the $m$-th BS antenna element is given by $d_{m}=\|{\bf p}_m - {\bf q}\|_2$.

Given the large aperture of the array adopted by the BS, we consider a practical scenario where the UE is within the near field region of the BS.
Therefore, and based on the above geometry, the channel coefficient between the UE and the $m$-th BS antenna at the $k$-th subcarrier is given by
\begin{equation}\label{ch-coeff}
  [{\bf h}_k]_m = \frac{\rho_k\lambda_k}{4\pi d_m} e^{-j \frac{2\pi}{\lambda_k} d_{m}}, ~ \forall k=1,\dots,K, ~ m=1,\dots,M,
\end{equation}
where $\lambda_k$ is the wavelength of the $k$-th subcarrier and $\rho_k$ subsumes all the other factors such as antenna pattern of the BS array, pathloss, and channel gain at the $k$-th subcarrier.

\section{Problem Formulation} \label{sec:Prob}

Given the system and channel models described in \sref{sec:System}, we cast the problem as a joint design of phase shifts of the quantized phase shifters and the delay configurations of the TD units.
{
The overall objective is to maximize the average beamforming gain, or equivalently, the signal-to-noise ratio (SNR), of the system across all the subcarriers.
}
Formally, this objective can be written as
\begin{align}\label{prob}
 \max\limits_{\boldsymbol{\tau}, \boldsymbol{\theta}} \hspace{2pt} & \hspace{2pt}  \frac{1}{K}\sum_{k=1}^{K}\left|\mathbf{w}_k^H\mathbf{h}_k\right|, \\
 \subto  \hspace{2pt} & \hspace{2pt} \tau_n\in \left[0, \tau_{\max}\right], ~ \forall n=1,2,\dots,N, \label{cons-1} \\
 & \hspace{2pt} \theta_m\in\boldsymbol{\Psi}, ~ \forall m=1,2,\dots,M, \label{cons-2}
\end{align}
where $\mathbf{w}_k$ is given by \eqref{bf-vec}, $\tau_{\max}\in\mathbb{R}_+$ denotes the maximum delay that the TD unit supports, and $\boldsymbol{\Psi}$ is a finite set with $2^r$ possible phase values drawn uniformly from $(-\pi, \pi]$.
In addition to the constraints \eqref{cons-1} and \eqref{cons-2}, we also assume that the channels, $\mathbf{h}_k, \forall k$, are unknown, as acquiring the channel state information (CSI) is typically challenging in systems with fully analog transceiver architectures \cite{Alkhateeb2014}.


\section{Proposed Solution}

In this section, we present our proposed near field wideband beam focusing solution.
The proposed approach decomposes the original joint design problem \eqref{prob} into two subproblems:
(i) The quantized analog phase shifter design subproblem, which focuses on {maximizing the beamforming gain at the center frequency}, while setting all the TD units to have zero delays;
and (ii) the TD unit design subproblem, which focuses on {maximizing the average beamforming gain achieved across all the subcarriers} by properly configuring the TD units.
Despite this decomposition, we will show in \sref{sec:simu} that the performance achieved by the proposed solution approaches that of the prior work \cite{Cui2021Near-Field} which (different than our solution) requires full channel knowledge and uniform array geometry. The motivation of this decomposition is partially originating from the observation that unless the number of elements in each sub-array is very large, then the performance degradation caused by the wideband effect for this sub-array is negligible.

\subsection{Decomposition of the Original Problem}
The combining vector in \eqref{bf-vec} can be expressed as $\mathbf{w}_k=\mathbf{w}_{\mathrm{TD},k}\odot\mathbf{w}_{\mathrm{PS}}$, where $\mathbf{w}_{\mathrm{TD},k}=e^{-j2\pi f_k\boldsymbol{\tau}\otimes\mathbf{1}_P}$ is the TD-based frequency-dependent combining vector and  $\mathbf{w}_{\mathrm{PS}}=\frac{1}{\sqrt{M}}e^{j\boldsymbol{\theta}}$ is the phase-shifter based frequency-flat combining vector. The symbol $\odot$ denotes the element-wise product.
Correspondingly, the design of the quantized analog phase shifters can be formulated as the following optimization problem
\begin{align}\label{prob-ps}
 \boldsymbol{\theta}^{\star} = \mathop{\arg\max}\limits_{\boldsymbol{\theta}} \hspace{2pt} & \hspace{2pt}  \left|\mathbf{w}_{\mathrm{PS}}^H\mathbf{h}_{f_c}\right|, \\
  \subto  \hspace{4pt} & \hspace{5pt} [\mathbf{w}_{\mathrm{PS}}]_m = \frac{1}{\sqrt{M}}e^{j\theta_m}, \label{ps-cons-1} \\
           & \hspace{5pt} \theta_m\in\boldsymbol{\Psi}, ~ \forall m=1,2,\dots,M, \label{ps-cons-2}
\end{align}
where $\mathbf{h}_{f_c}$ denotes the channel at the center frequency. After obtaining $\boldsymbol{\theta}^{\star}$, the design problem of the TD units can be expressed as follows
{
\begin{align}\label{prob-td}
 \boldsymbol{\tau}^{\star} = \mathop{\arg\max}\limits_{\boldsymbol{\tau}} \hspace{2pt} & \hspace{2pt}  \frac{1}{K}\sum_{k=1}^{K}\left|\left(\mathbf{w}_{\mathrm{TD},k}\odot\widehat{\mathbf{w}}_{\mathrm{PS}}^\star\right)^H\mathbf{h}_k\right|, \\
 \subto  \hspace{2pt} & \mathbf{w}_{\mathrm{TD},k}=e^{-j2\pi f_k\boldsymbol{\tau}\otimes\mathbf{1}_P}, \label{td-cons-1} \\
 & \hspace{2pt} \tau_n\in \left[0, \tau_{\max}\right], ~ \forall n=1,2,\dots,N, \label{td-cons-2} \\
 & \hspace{2pt} [\widehat{\boldsymbol{\theta}}^\star]_m = \argmin_{\theta\in\boldsymbol{\Psi}}\left|\theta - \left([\boldsymbol{\theta}^\star]_m + 2\pi f_c\tau_n\right)\right|, \notag \\
 & \hspace{2pt} \forall n, \forall m\in\{(n-1)P+1, \dots, nP\},  \label{td-cons-3}
\end{align}
where $\widehat{\mathbf{w}}_{\mathrm{PS}}^\star=\frac{1}{\sqrt{M}}e^{j\widehat{\boldsymbol{\theta}}^{\star}}$. The constraint \eqref{td-cons-3} is to ensure that the gain at the center frequency, i.e., the objective function of \eqref{prob-ps}, will not be influenced after the introduction of the TD units.
}
Next, in Sections \ref{subsec:PS} and \ref{subsec:TD}, we present the proposed solutions for solving \eqref{prob-ps} and \eqref{prob-td}.

\subsection{The Design of the Analog Phase Shifters} \label{subsec:PS}

\begin{figure}[t]
	\centering
	\includegraphics[width=1\columnwidth]{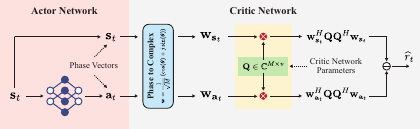}
	\caption{The adopted actor-critic learning architecture where the actor network is implemented using the fully-connected neural network and the critic network is designed to reflect the underlying signal model.}
	\label{model-critic}
\end{figure}

As mentioned before, the design of the quantized analog PSs focuses on maximizing the beamforming/combining gain \textbf{at the center frequency $f_c$}.
We adopt the similar reinforcement learning (RL) formulation as in the previous work \cite{Zhang2022Reinforcement}.
Specifically, we define the \textbf{state} ${\bf s}_t$ as a vector that consists of the phases of all the phase shifters at the $t$-th iteration, that is, ${\bf s}_t=\left[\theta_1, \theta_2, \dots, \theta_M\right]^T$.
We define the \textbf{action} ${\bf a}_t$ as the element-wise changes to all the phases in ${\bf s}_t$. Since the phases can only take values in $\boldsymbol\Psi$, a change of a phase represents the action that a phase shifter selects a value from $\boldsymbol\Psi$. Therefore, the action is directly specified as the next state, i.e., ${\bf a}_t = {\bf s}_{t+1}$, which can be viewed as a deterministic transition in the Markov Decision Process (MDP).
We define the \textbf{reward} as the achieved beamforming gain difference between the action (i.e., new beam, denoted as $\mathbf{w}_{\mathrm{PS}, t}$) and the state (i.e., the previous beam, denoted as $\mathbf{w}_{\mathrm{PS}, t-1}$). Formally, $r_t=|\mathbf{w}_{\mathrm{PS}, t}^H\mathbf{h}_{f_c}|^2 - |\mathbf{w}_{\mathrm{PS}, t-1}^H\mathbf{h}_{f_c}|^2$.

Moreover, to improve the sample efficiency of the proposed algorithm in learning over large antenna arrays, we develop a signal model inspired online RL-based search algorithm.
Specifically, we improve the existing actor-critic architecture based beam learning algorithm \cite{Zhang2022Reinforcement} by designing a special critic network that better utilizes the underlying signal model.
Formally, the critic network takes the following form
\begin{equation}\label{critic}
  f({\bf s}_t, {\bf a}_t) = \mathbf{w}_{{\bf a}_t}^H\mathbf{Q}\mathbf{Q}^H\mathbf{w}_{{\bf a}_t} - \mathbf{w}_{{\bf s}_t}^H\mathbf{Q}\mathbf{Q}^H\mathbf{w}_{{\bf s}_t},
\end{equation}
where $\mathbf{Q}\in\mathbb{C}^{M\times v}$ is the model parameters with $v\in\mathbb{N}_+$ being a hyperparameter, $\mathbf{w}_{{\bf a}_t}=\frac{1}{\sqrt{M}}e^{j{\bf a}_t}$ and $\mathbf{w}_{{\bf s}_t}=\frac{1}{\sqrt{M}}e^{j{\bf s}_t}$.
The complete beam learning framework is illustrated in \fref{model-critic}.

\subsection{The Design of the Time Delay Units} \label{subsec:TD}

In this subsection, we introduce the proposed low complexity geometry-assisted TD unit design algorithm.
Before we delve into the details, it is worth mentioning that the motivation of leveraging geometry to design the delay configuration of the TD units can be found in the following observations.
\textbf{First,} the TD units are mainly adopted to compensate for the different delays experienced at the different parts of the large array. This implies that the design can be determined based on the \textit{distance differences} between the elements on the antenna array and the UE.
\textbf{Second,} the absolute distance difference between any two antenna elements on the array and the UE is upper bounded by the array aperture, i.e., $D$.
\textbf{Third,} depending on the relative positions of the BS and UE, the delay profile on the BS's array (i.e., the experienced delay value as a function of antenna element) presents different monotonicity properties.

These observations provide useful insights and guidance for the considered TD units design as they help constrain the search space of the delay configurations.
{
Moreover, as indicated by the first observation, the introduction of the TD units is to compensate for the differences in the propagation delays experienced at the different receive antenna elements, which are directly related to the different propagation distances. Therefore, understanding and finding such distance differences could facilitate the design of the TD units.
}
Based on this, we introduce the concept of the \textit{distance difference function} (DDF), which directly determines the design of the delay units.
To be more specific, we set the point ${\bf p}_{\mathrm{ref}}=[0, \frac{D}{2}]^T$ as the \emph{reference point} of the BS array.
Further, we define $\Delta = 1-\alpha$ as the \emph{relative coefficient} with respect to the \emph{reference point}\footnote{As described in \sref{subsec:H}, the position of an antenna is determined via its coefficient $-1\le \alpha\le 1$.}, which implies that $\Delta\in[0, 2]$.
Therefore, the \emph{reference distance}, defined as the distance between ${\bf p}_{\mathrm{ref}}$ and the considered UE position ${\bf q}$, is given by $d_{\mathrm{ref}}=\|{\bf p}_{\mathrm{ref}} - {\bf q}\|_2$.
Based on the reference distance, we define the DDF $f_{\mathrm{d}}$ for any point on the array $\bp_\Delta=[0, (1-\Delta)\frac{D}{2}]^T$ and the UE as
\begin{equation}\label{ddf}
  f_{\mathrm{d}}(\Delta, {\bf q}) = d(\Delta) - d_{\mathrm{ref}},
\end{equation}
where $d(\Delta) = \|\bp_\Delta - {\bf q}\|_2$.
As can be seen, the DDF depends only on $\Delta$ and ${\bf q}$. Besides, $d_{\mathrm{ref}}$ corresponds to $\Delta=0$, and $f_{\mathrm{d}}=0$ when $\Delta=0$.
Moreover, an important observation is that the monotonicity property of $f_{\mathrm{d}}$ depends on the UE position ${\bf q}$ relative to the BS array.
Without loss of generality, we assume $q_x>0$.
Then, the UE position space can be divided into three regimes based on the value of $q_y$: (i) $q_y>\frac{D}{2}$; (ii) $q_y<-\frac{D}{2}$; and (iii) $|q_y|<\frac{D}{2}$.
For these regimes, $f_{\mathrm{d}}$ is monotonically increasing  when $q_y>\frac{D}{2}$ and is monotonically decreasing when $q_y<-\frac{D}{2}$.
When $|q_y|<\frac{D}{2}$, $f_{\mathrm{d}}$ first decreases and then increases.
{
It is worth noting that the delay configuration of the TD units is closely related to $f_{\mathrm{d}}$ which essentially characterizes the additional propagation distance at different parts of the array.
Specifically, the ideal delay configuration of the $m$-th antenna is given by $\frac{f_{\mathrm{d}}(\Delta_m, {\bf q})}{c}$, where $\Delta_m=1-\alpha_m$.
However, as can be seen from \eqref{ddf}, in order to obtain the exact expression of $f_{\mathrm{d}}$, the UE position ${\bf q}$ should be known. Since in practice systems, however, the UE position is either  unavailable or noisy, it is important to develop search algorithms that are able to find an approximation of $f_{\mathrm{d}}$ , and hence design the delay configurations, \textbf{without requiring the UE position.}
Next, we describe the proposed low complexity delay search method. Each search cycle in this method consists of the following three steps:
}

\subsubsection{Select a Linear Approximation}
Inspired by the behavior of the DDF $f_{\mathrm{d}}$, we propose a low complexity linear approximation based search algorithm to design the delay configurations of the TD units.
Specifically, we use the following piecewise linear function to approximate $f_{\mathrm{d}}$, that is
\begin{equation}\label{linear-approx}
\widehat{f}_{\mathrm{d}}(\Delta; \{a_x,a_y,b\}) = \left\{
\begin{array}{ll}
\frac{a_y}{a_x}\Delta, & 0 \le \Delta \le a_x, \\
\frac{b-a_y}{2-a_x}(\Delta - a_x) + a_y, & a_x < \Delta \le 2,
\end{array}
\right.
\end{equation}
where the approximation function $\widehat{f}_{\mathrm{d}}$ is parameterized by $a_x,a_y,b$,
{
with $(a_x, a_y)$ modelling the coordinate of the only breakpoint of \eqref{linear-approx} and $b$ modelling the intersection of the second part of \eqref{linear-approx} with $\Delta=2$.
}
Therefore, the approximation problem essentially becomes a joint search problem of $a_x$, $a_y$ and $b$.
Since $a_x$, $a_y$ and $b$ should satisfy the conditions determined by the properties of $f_{\mathrm{d}}$,
$0\le a_x\le 2$, $-\frac{D}{2}a_x\le a_y\le\frac{D}{2}a_x$, $-D\le b\le D$, we discretize the original continuous search space into a finite size discrete grid.
Specifically, we denote $\mathcal{A}_x\subseteq[0, 2]$ as the finite search set that contains the search points of $a_x$,  $\mathcal{A}_{y|a_x}\subseteq[-\frac{D}{2}a_x, \frac{D}{2}a_x]$ as the finite search set of $a_y$ given $a_x$, and $\mathcal{B}\subseteq[-D, D]$ as the finite search set of $b$. In each search cycle, one combination of the possible values of $a_x$, $a_y$, and $b$ are selected, which gives one candidate approximation of $f_{\mathrm{d}}$.

\subsubsection{Compute the Delay Configurations}

After obtaining $\widehat{f}_{\mathrm{d}}$, the delay search is then conducted based on sampling of the corresponding \emph{delay difference function}, defined as $\widehat{f}_{\tau}=\frac{\widehat{f}_{\mathrm{d}}}{c}$.
If we set one sampling point at the center of each sub-array to have the set $\{\Delta_1, \Delta_2, \dots, \Delta_N\}$, with $0 \le \Delta_1 < \Delta_2 < \cdots < \Delta_N \le 2$,
then, the delay configuration vector will be
\begin{equation}\label{TD-config}
  \boldsymbol{\tau}=\left[\widehat{f}_{\tau}(\Delta_1), \widehat{f}_{\tau}(\Delta_2), \dots, \widehat{f}_{\tau}(\Delta_N)\right]^T.
\end{equation}

{
\subsubsection{Evaluate the Configured Delays}
In this step, the analog phase shifters are first reconfigured to compensate for the additional phase shifts introduced by the TD units, according to \eqref{td-cons-3}.
Then, the system measures the receive beamforming gain at the different subcarriers to determine the performance of the current delay configurations. This completes one cycle of the proposed delay search algorithm.
}

\subsection{Complexity Analysis}

For the analog PS design problem, the proposed signal model-based learning architecture reduces the computational complexity when comparing with the signal model-unaware architecture \cite{Zhang2022Reinforcement}.
Specifically, the complexity reduction of the proposed solution mainly comes from the following aspects.
\textbf{First,} the signal model-based critic network design reduces its complexity from $\mathcal{O}(M^2N_{\mathrm{b}}N_{\mathrm{iter}})$, where we assume a fully-connected neural network-based critic network architecture as in \cite{Zhang2022Reinforcement}, to $\mathcal{O}(MN_{\mathrm{b}}N_{\mathrm{iter}})$, with $N_{\mathrm{b}}$ denoting the mini-batch size and $N_{\mathrm{iter}}$ denoting the total number of iterations.
\textbf{Second,} from training perspective, the convergence of the proposed solution is much faster than the signal model-unaware architectures, meaning that $N_{\mathrm{iter}}$ is normally also small.
To be more specific, from the simulation result, it indicates that, for a $256$-antenna BS with $3$-bit quantized PSs, the training of the critic network, depending on the number of real power measurements, is able to converge within $1000\sim5000$ iterations. And with a well-trained critic network, the actor network is typically able to converge within only $20\sim50$ iterations.
\textbf{Finally,} due to the decoupling of the original joint TD-PS design problem \eqref{prob}, the search space becomes much smaller, which generally leads to less search complexity.

\section{Simulation Results} \label{sec:simu}

In this section, we evaluate the performance of the proposed near field wideband beamforming design algorithm.

\subsection{Simulation Setup}

In this simulation, we consider the scenario where a BS receiver adopts a linear array with $256$ elements.
While the positions of the antenna elements within the array are randomly sampled, giving rise to an unknown and non-uniform array geometry, we assume that the aperture of the array is given by $(M-1)\frac{\lambda_c}{2}$, with $\lambda_c$ being the wavelength of the center frequency and $M=256$.
Furthermore, we assume that the system is operating at a center frequency of $100$ GHz (hence $\lambda_c=0.3$ centimeter) and has {a bandwidth of $10$ GHz}.
Moreover, each antenna of the array is followed by a $3$-bit analog PS.
The signals from the $P$ adjacent PSs are first combined and then processed by the same TD unit, as illustrated in \fref{sys-ps-ttd}.
The value of $P$ depends on the number of TD units in the specific simulations.

\subsection{Numerical Results}

\begin{figure}[t]
	\centering
    \subfigure[Beam focusing at different frequencies]{ \includegraphics[width=0.46\columnwidth]{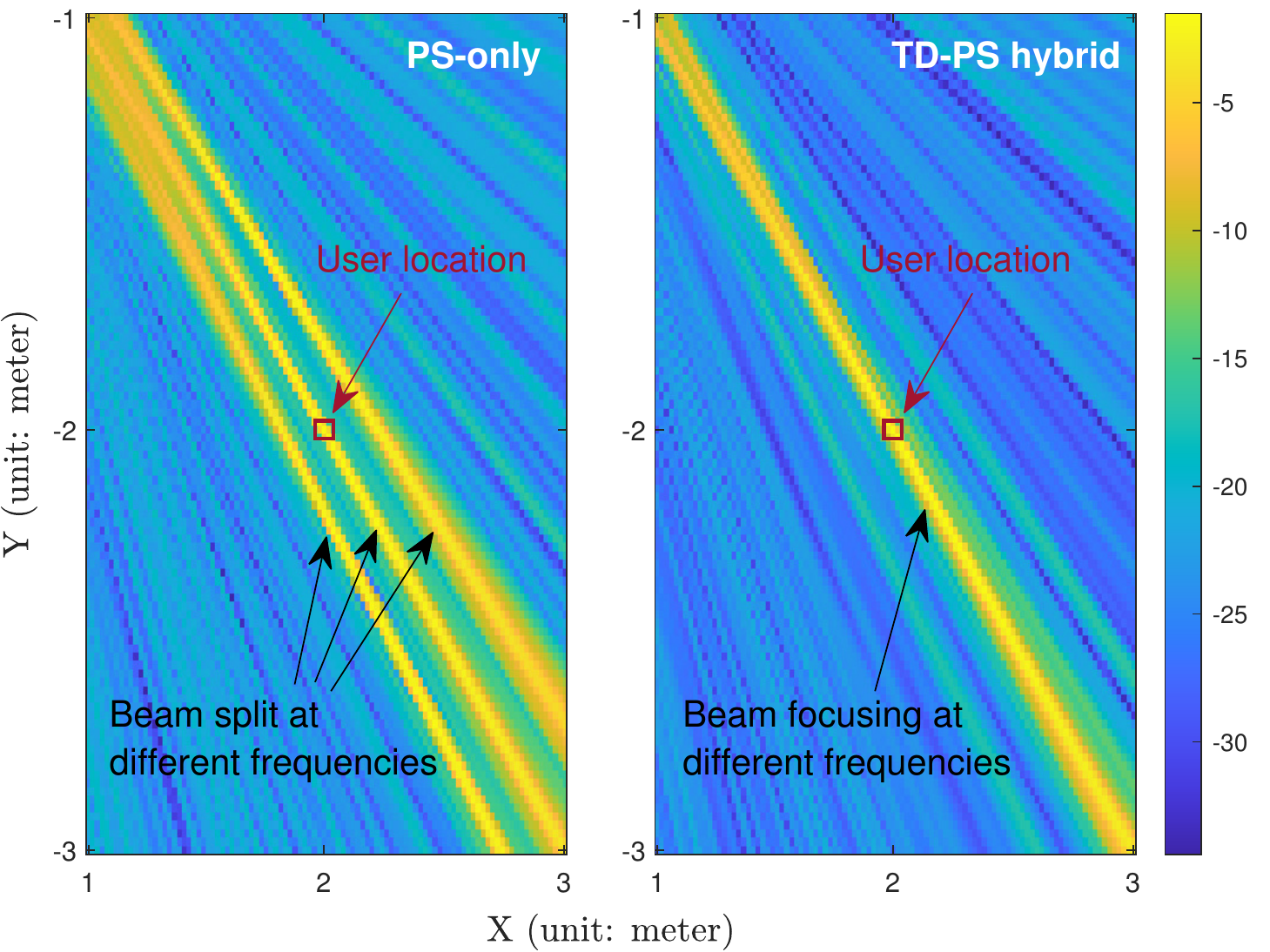} \label{fig:focusing} }
    \subfigure[Wideband performance]{ \includegraphics[width=0.43\columnwidth]{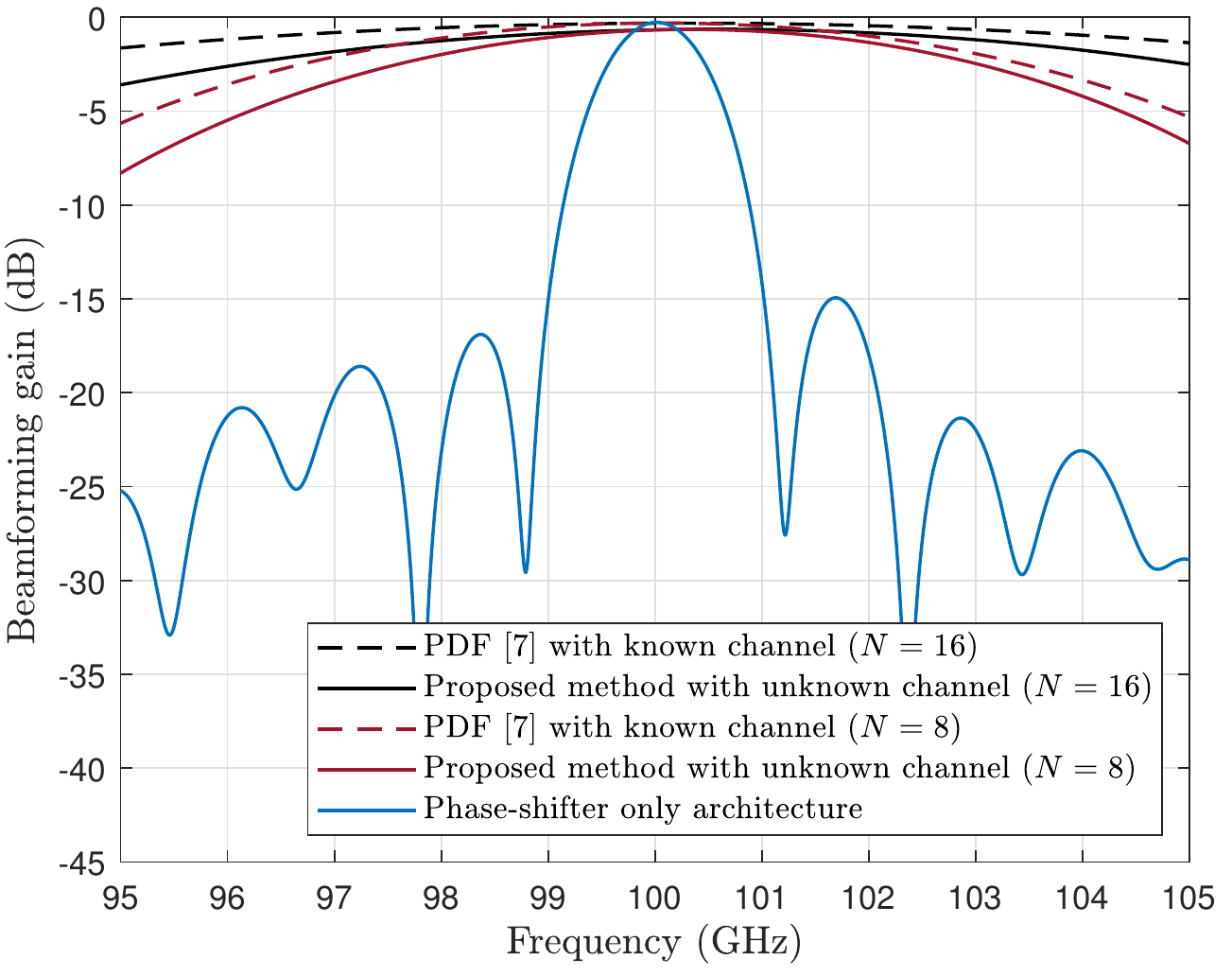} \label{fig:wideband} }
	\caption{The evaluation of the proposed solution. (a) shows the beam focusing performance at different frequencies (lowest, center and highest frequencies of the considered bandwidth) to a user at ${\bf q}=[2, -2]^T$ that is marked by the red square, where the figure to the left shows the performance of PS-only architecture and the figure to the right shows the performance of the TD-PS hybrid architecture with the proposed beam focusing algorithm. (b) shows the beamforming gain performance with different numbers of TD units.}
	\label{fig:simu}
\end{figure}

In \fref{fig:focusing}, we visualize the beam focusing performance of the proposed solution at different frequencies to a user at ${\bf q}=[2, -2]^T$, where we show the lowest, center and highest frequencies of the considered bandwidth.
The average beamforming gain performance at these frequencies is plotted on the considered user grid.
Specifically, the figure to the left in \fref{fig:focusing} shows the performance of the PS-only architecture with the phase shifts designed based on the center frequency.
As can be seen, due to the frequency-flatness of the analog phase-shifters, the beam can only focus on the considered user at the center frequency, while being completely deviated at the other two frequencies, an effect that is oftentimes referred to as beam split.
By leveraging the TD-PS hybrid architecture and the proposed beam focusing solution, the beamforming vector becomes frequency-dependent. As a result, \textbf{the beams at different frequencies are all able to focus on the user,} as shown in the figure to the right in \fref{fig:focusing}.

The wideband performance of the proposed algorithm when utilizing different numbers of TD units is then evaluated.
As can be seen from \fref{fig:wideband}, if the BS adopts a PS-only architecture (without TD units), the system experiences quite severe performance degradation in the considered bandwidth.
Specifically, although the system has {a bandwidth of $10$ GHz}, the 3 dB bandwidth achieved by the PS-only architecture is only around $1$ GHz.
Moreover, most of the subcarriers experience less than $-10$ dB beamforming gain with considerable subcarriers having even over $-20$ dB beamforming gain loss compared to the center frequency.
However, \textbf{by leveraging only $8$ TD units, meaning that every $32$ antennas share one TD unit, the performance gets improved quite significantly, where the 3 dB bandwidth is widened to almost $6$ GHz.}
If the number of the TD units is allowed to be further increased to 16 TD units, the system is able to achieve almost frequency-flat performance.
To better understand the achieved performance by the proposed solution, we compare it with the prior work \cite{Cui2021Near-Field}, where the authors propose a phase-delay focusing (PDF) method that achieves beam focusing with perfect channel knowledge.
As can be seen in \fref{fig:wideband}, the proposed solution is able to approach the PDF performance with negligible difference.
It is worth noting that \textbf{the proposed solution achieves such performance while not requiring any channel knowledge and while being generic to arrays with arbitrary geometries.}
This reduces the system operation overhead and enhances the robustness of the beam focusing solution against various hardware imperfections.

\section{Conclusions} \label{sec:Con}

In this paper, we developed a low-complexity wideband beamforming solution for hybrid TD-PS based RF architectures operating in the near-field regime.
To perform beamforming, the proposed solution does not require any explicit channel knowledge and relies only on the power measurements of the received signal at different subcarriers.
The sample efficiency of the proposed solution is enhanced by designing a signal model inspired actor-critic architecture to learn the phase shifts, and by leveraging a geometry-assisted linear approximation model to configure the TD units.
Important future extensions of this work include near-field wideband beam codebook learning, interference aware near-field beam pattern learning, and beam/codebook learning under user mobility.


\end{document}